\documentclass[twocolumn,aps,amsmath,amssymb,prl,epsf,showpacs]{revtex4}
\usepackage{psfig}
\usepackage{graphicx}% Include figure files
\usepackage{dcolumn}% Align table columns on decimal point
\usepackage{bm}% bold math
\newcommand{\be}{\begin{equation}}
\newcommand{\ee}{\end{equation}}
\def\bq{\begin{eqnarray}}
\def\eq{\end{eqnarray}}
\def\beq{\begin{eqnarray}}
\def\eeq{\end{eqnarray}}

\def\ba{\begin{eqnarray}}
\def\ea{\end{eqnarray}}

\newcommand{\mpl}{m_{\rm p}}
\newcommand{\lpl}{\ell_{\rm p}}

\newcommand{\M}{\ensuremath{{ M}}}
\newcommand{\pd}{\ensuremath{{\dot{\phi}}}}
\newcommand{\pdd}{\ensuremath{{\ddot{\phi}}}}
\newcommand{\al}{\ensuremath{\alpha}}
\begin{document}
\preprint{AEI--2005--020}
\preprint{IGPG-05/3-3}
\title{A black hole mass threshold from non-singular
quantum gravitational collapse}
%%%%%%%%%%%%%%%%%%%%%%%%%%%%%%%%%%%%%%%%%%%%%%%%%%%%%%%%%%%%%%%%%%%%%%

\author{Martin Bojowald$^1$, Rituparno Goswami$^2$, Roy
Maartens$^3$, Parampreet Singh$^4$}

\affiliation{$^1$Max-Planck-Institut f\"ur Gravitationsphysik,
Albert-Einstein-Institut, D-14476 Potsdam, Germany}

\affiliation{$^2$Tata Institute for Fundamental Research, Colaba,
Mumbai, India}

\affiliation{$^3$Institute of Cosmology and Gravitation,
University of Portsmouth, Portsmouth PO1~2EG, UK}

\affiliation{$^4$Institute for Gravitational Physics and Geometry,
Pennsylvania State University, PA 16802, USA}

\date{\today}
%%%%%%%%%%%%%%%%%%%%%%%%%%%%%%%%%%%%%%%%%%%%%%%%%%%%%%%%%%%%%%%%%%%%%%

\begin{abstract}

Quantum gravity is expected to remove the classical singularity that
arises as the end-state of gravitational collapse. To investigate
this, we work with a toy model of a collapsing homogeneous scalar
field. We show that non-perturbative semi-classical effects of Loop
Quantum Gravity cause a bounce and remove the black hole
singularity. Furthermore, we find a critical threshold scale, below
which no horizon forms -- quantum gravity may exclude very small
astrophysical black holes.

\end{abstract}


\maketitle
%%%%%%%%%%%%%%%%%%%%%%%%%%%%%%%%%%%%%%%%%%%%%%%%%%%%%%%%%%%%%%%%%%%%%%

Singularity formation during gravitational collapse signals the
breakdown of classical general relativity. In a more complete
theory of quantum gravity the singularity should be removed.
However, a satisfactory quantum gravity theory has yet to be
developed. In addition, the dynamics of general collapse is very
complicated. Thus we can only expect to make partial progress in
tackling the problem, using a candidate for quantum gravity
and a collapse model that is simple enough to be tractable.

A non-perturbative approach to quantizing gravity is loop quantum
gravity or quantum geometry~\cite{loopqg}, which gives rise to a
discrete spatial structure \cite{AreaVol} and whose successes include
prediction of black hole entropy~\cite{ABCK:LoopEntro}. Applied to the
early universe, loop quantum effects can remove the big bang
singularity~\cite{Sing}. A natural question is: do these effects also
remove the black hole singularity as the end-state of collapse? 
Techniques to handle inhomogeneous systems are under development
and give promising indications \cite{InHom}, but
they do not easily reveal the physical picture.
We thus consider a simple toy model of a
collapsing {\em homogeneous} scalar field. Classically, this model always
produces a black hole, but we show that loop quantum effects change
this situation dramatically.

Since we do not yet know semi-classical non-perturbative effects in
inhomogeneous cases, we are unable to perform our analysis in the
general case.  However, when we split the system into a homogeneous
star interior and an inhomogeneous outside region, known quantum
effects in the interior can be carried into the exterior indirectly
through matching conditions.  The collapsing homogeneous scalar field
cannot be matched to a Schwarzschild exterior because the pressure
does not vanish at the boundary. But in any case, we expect that
quantum effects will include
%>>>new:
small
%<<<
non-stationary corrections
%>>>delete:
% (which should be negligible for large black holes). 
%So it is reasonable to match the interior to 
%<<<
%>>>new:
and thus use 
%<<<
a non-stationary spherically symmetric exterior.  The generalized
Vaidya metric provides a reasonable starting point. It is sufficiently
general to allow for a broad range of behavior, including
%>>>delete: non-stationary 
radiative effects.

Our analysis is based on effective equations for the
interior which have been established in the cosmological
setting. Fundamentally, the evolution is described by a wave function
subject to a difference equation, and effective equations describe the
motion of semiclassical wave packets \cite{time}. As long as one stays
in semiclassical regimes, which can e.g.\ be checked using the size of
curvature, one gets reliable expectations for the quantum situation.

We first review the classical collapse and the inevitability
%>>>delete: existence 
of a black hole singularity covered by a horizon, for any initial
mass. The
%>>>new: 
isotropic
%<<<
interior metric is
%>>>new:
\cite{HawkingEllis}
%<<<
\begin{equation}
ds^2=-dt^2+a(t)^2(1+r^2/4)^{-2} \left[dr^2+r^2d\Omega^2\right]\,,
\label{eq:metric}
\end{equation}
and the massless scalar field $\phi(t)$ has pressure and energy
density $p=\rho=\frac{1}{2}\pd^2$.  The Friedmann equation is
\begin{equation}
{\dot{a}^2 / a^2}= {4\pi\lpl^2 }\,{\dot{\phi}^2/3 }-{1/ a^2}\,.
\label{eq:ein1}
\end{equation}
The Klein-Gordon equation, $a\pdd+3\dot a\,\pd=0$, has solution
\begin{equation}
\pd={L}/{a^3}\,, \label{eq:pd1}
\end{equation}
where $L$ is a length scale associated with the maximal size of
the collapse region, since (\ref{eq:ein1})
implies
\begin{equation}\label{am}
a\leq a_{\rm m}\equiv \left({4\pi/3} \right)^{1/4} \sqrt{\lpl
L}\,.
\end{equation}
At the singularity $a\rightarrow 0$, we have
$\pd,\rho\,\rightarrow\infty$. The solution of the Friedmann
equation is
\begin{equation}
t-t_0=a_{\rm m}\int_{a/a_{\rm m}}^{a_0/a_{\rm m}}\, {b^2db \over
\sqrt{1-b^4}}\,, \label{eq:a1}
\end{equation}
where $a_0(\leq a_{\rm m})$ gives the initial size of the collapse
region at time $t_0$. The singularity $a=0$ is covered
by a horizon (see below), and reached in finite proper time
for any $a_0$:
\begin{equation}
{1\over a_{\rm m}}(t_{\rm s}-t_0)<\int_0^1{db \over
\sqrt{1-b^4}}={1\over \sqrt{2}}F\left({\pi\over2},{1\over\sqrt
2}\right),
\end{equation}
where $F$ is an elliptic integral of the first kind.

%>>>reformulation:
We now add non-perturbative modifications to the dynamics, motivated by
loop quantum gravity \cite{dyn}.  The quantization introduces a
%<<<
fundamental length scale
\begin{equation}\label{ls}
\ell_*=0.28\sqrt{ j }\,\lpl\,,
\end{equation}
where $j (>1)$ is a half-integer that is freely specifiable. For
$a<\ell_*$, the dynamics is increasingly different from general
relativity. For $a\lesssim \lpl$, the continuum approximation to the
spacetime geometry begins to break down, and the fully quantum gravity
regime is reached. In the intermediate regime $\lpl
\lesssim a \lesssim \ell_*$, loop quantum effects may be treated
semi-classically, i.e., the spacetime metric behaves classically,
while the dynamics acquires non-perturbative modifications to
general relativity~\cite{time}. The non-perturbative
semi-classical regime exists provided $\ell_*\gg \lpl$, i.e., for
$j\gg1$.

The key feature of the loop quantization scheme is the prediction
that the geometrical density, $1/a^3$, does not diverge as $a \to
0$, but remains finite. The expectation values of the density
operator are approximated by
%\begin{equation}
$d_{j}(a)=  D(a)\, a^{-3}$,
%\end{equation}
where the loop quantum correction factor is~\cite{Ambig}
\begin{eqnarray}
&&D(a) = \left( {8/ 77}\right)^6 q^{3/2} \Big\{7 \Big[(q+1)^{11/4}
-|q-1|^{11/4}\Big] \nonumber \\
&&~~~{}- 11q\Big[(q+1)^{7/4}-{\rm sgn}\,(q-1) |q-1|^{7/4}\Big]
\Big\}^6\!,\label{D}
\end{eqnarray}
with $q\equiv {a^2 / \ell_*^2}$. In the classical limit we recover
the expected behavior of the density, while the quantum regime
shows a radical departure from classical behavior:
\begin{eqnarray}\label{dq}
a \gg \ell_*:  D \approx 1\quad,\quad a \ll \ell_* :
D \approx \left({12/7}\right)^6\left({a/\ell_*}\right)^{15}\!.
\end{eqnarray}
Then $d_j$ remains finite as $a \to 0$, unlike in conventional quantum
cosmology, thus evading the problem of the big-bang singularity in a
closed model~\cite{BounceClosed}. Intuitively, one can think of the
modified behavior as meaning that gravity, which is classically always
attractive, becomes repulsive at small scales when quantized. This
effect can produce a bounce where classically there would be a
singularity, and can also provide a new mechanism for high-energy
inflationary acceleration~\cite{Inflation}. In the semi-classical
regime (where the spectrum can be treated as continuous), $d_j$ has a
smooth transition from classical to quantum behavior, varying from
$a^{-3}$ to $a^{12}$.  We emphasize that this is but one possibility
for a bounce which we use for concreteness, while bounces in general
appear more generically in loop cosmology \cite{GenericBounce}.

In loop cosmology the Hamiltonian of a scalar field in a
closed universe is
\begin{equation}\label{hamiltonian}
{\cal H}= a^3V(\phi)+d_j \,{P_\phi^2}/2 \,, ~~  P_\phi =
d_j^{-1}\dot \phi\,,
\end{equation}
where $P_\phi$ is the momentum canonically conjugate to $\phi$.
This leads to a modified Friedmann equation~\cite{Inflation,Semi},
\begin{equation} \label{back}
{\dot{a}^2 \over a^2} = {8\pi \lpl^2\over 3}\left[V(\phi) +
{1\over 2D} \dot{\phi}^2 \right]-{1\over a^2}\,,
\end{equation}
and a modified Klein-Gordon equation~\cite{Closed}
\begin{equation}
\ddot \phi + 3\, a^{-1}\dot a \, (1 - \alpha) \, \dot \phi
+DV(\phi)=0 \, , ~~ \alpha  \equiv a \dot D/(3 \dot a D)\,.
\label{kgq}
\end{equation}
For $a \ll \ell_*$, we have $\alpha \to 5$, whereas classically $D
= 1$ and hence $\alpha = 0$. Thus in the semi-classical regime,
$0<\alpha \leq5$.

For a massless scalar field, $V=0$, the solution of
Eq.~(\ref{kgq}), generalizing Eq.~(\ref{eq:pd1}), is
\begin{equation}\label{phi}
\dot\phi=Ld_j(a)\,,
\end{equation}
so that $P_\phi=L=\,$const. Then the Friedmann equation becomes
\begin{equation}\label{fq}
\dot{a}^2+1=D(a)(a_{\rm m}/a)^4\,.
\end{equation}
The energy density and pressure are modified as $\rho = \dot
\phi^2/2D\,,$  $p = \dot\phi^2\, \left(1 - \alpha \right)/2D\,,$
so that
\begin{equation}
 w\equiv p/\rho=1-\alpha \,. \label{peq}
\end{equation}
(The modified $\rho$ and $p$ satisfy the usual conservation
equation if $\phi$ satisfies the modified Klein-Gordon equation.)
Since $\alpha$ varies from 0 to 5 as $a$ decreases, the $\dot
\phi$ term in Eq.~(\ref{kgq}), which classically behaves as {\em
anti-frictional} during collapse, starts to behave as {\em
frictional} when $\alpha > 1$. Thus, contrary to classical
behavior, where $\dot \phi$ increases as $a$ decreases, in the
semi-classical regime the scalar field starts slowing down with
collapse. In fact at $\alpha = 2$ the magnitude of the frictional
term becomes exactly equal to the classical anti-frictional term.
Thereafter at smaller values of the scale factor the term becomes
increasingly frictional and the collapse further slows down, and
may turn around.

\begin{figure}
 \includegraphics[width=9cm,height=6cm]{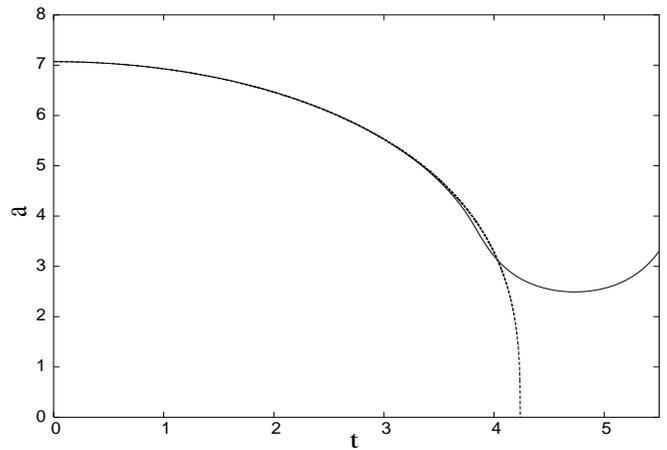}
 \caption{
The scale factor $a(t)$ of the collapsing interior, for classical
(dashed) and semi-classical quantum dynamics (solid).
\label{collapse_fig}}
\end{figure}

The point where $\alpha = 2$ is also the point beyond which the
null energy condition is violated: $w<-1$, by Eq.~(\ref{peq}).
Violations of the null energy condition by quantum gravity effects
are to be expected, and in loop quantum gravity this occurs for
$\alpha > 2$, when the scalar field effectively behaves as a
``phantom" field.

In order to see qualitatively how the non-perturbative frictional
quantum effects remove the classical singularity, we assume that,
over a small interval of scale factor, we can take $\alpha
\approx\,$constant, so that $D \approx D_* \,\left(
{a/\ell_*}\right)^{3\al}\,,$ where $D_*$ is a dimensionless
constant. By Eq.~(\ref{phi}),
\begin{equation}
\pd \approx LD_* \ell_*^{-3\alpha}\,a^{3(\alpha-1)}\,,
\label{eq:pd2}
\end{equation}
which shows how the kinetic energy decreases with decreasing $a$
when $\alpha>1$, contrary to the classical case. The modified
Friedmann equation~(\ref{fq}) gives
\begin{equation}
\dot{a}^2 \approx \left({a_{\rm m}^4 \ell_*^{-3\alpha}D_*}\right)
a^{3\alpha-4}-1\,.
\end{equation}
In general relativity, where $\alpha=0$ and $D_*=1$, this shows
that for $a<a_{\rm m}$, there is no turning point in $a$, i.e.,
$\dot a\not=0$. With loop quantum effects, for $\alpha>{4\over3}$,
the equation $\dot {a}(t_{\rm c})=0$ has a solution, $a_{\rm c}
\approx \left( {\ell_*^{3\alpha}/D_* a_{\rm m}^{4}}
\right)^{1/(3\alpha-4)} \ll a_{\rm m}\,.$ Thus the collapse leads
to a bounce and singularity avoidance. The numerical integration
of the modified Friedmann and Klein-Gordon equations confirms the
qualitative analysis, and the results are illustrated in
Fig.~\ref{collapse_fig}. As is clear from the figure, the
classical curve (dashed line) hits the singularity in finite time,
whereas the quantum-corrected curve bounces and avoids the
singularity. The key question is whether a horizon forms in the
quantum-corrected collapse.

The formation or avoidance of the singularity $a=0$ is independent
of the matching to the exterior. But in order to understand
horizon formation in the semi-classical quantum case, we need to
impose the matching conditions. Since the pressure is nonzero at
the boundary, given in comoving coordinates by $r=R=\,$constant,
the interior cannot be directly matched to a static Schwarzschild
exterior. However we can match to an intermediate non-stationary
region -- for example, a generalized Vaidya region~\cite{joshi},
\begin{equation}
ds^2=-\left[1-{2\M(v,\chi)}/{\chi} \right]dv^2
+2dvd\chi+\chi^2d\Omega^2\,. \label{eq:vaidya}
\end{equation}
The usual Vaidya mass $M/\lpl^2$ is generalized so that $\partial M
/\partial\chi $ may be nonzero. The total mass measured by an
asymptotic observer is $m=m_M+m_\phi$, where $m_M$ is the total mass
in the generalized Vaidya region, and $m_\phi=\int \rho dV$ the
interior mass. By Eqs.~(\ref{eq:metric}), (\ref{am}) and (\ref{phi}),
\begin{equation} \label{eq:mass}
{m_\phi \over \mpl}= {3a\over2\lpl}\left(\!{a_{\rm m}\over
a}\!\right)^{\!4}\!D(a)\!\left[\tan^{-1}{R\over2}-{R(1-R^2/4)\over
2(1+R^2/4)^2} \right]\!.
\end{equation}
Since we do not specify the matter content in the exterior, and
do not know the modified field equations there, we cannot
determine $M(v,\chi)$ and thus $m_M$. However, as we discuss
below, we can still draw qualitative conclusions about the
behavior of horizons close to the matter shells.

Matching the 1st and 2nd fundamental forms we obtain
\begin{eqnarray}
\chi(v)&=&Ra(t)/(1+R^2/4) \,,\label{eq:match1}\\
{dv}/{dt}&=& (1+R^2/4)/(1-R^2/4-R\dot{a})\,,
\label{eq:match2}\\
2M &=& aR^3(\dot{a}^2+1)/(1+R^2/4)^3\,, \label{eq:match3}\\
\label{Ktt} -\!M_{,v}&=&\chi_{,vv}\!+\!
(1-{2M}/{\chi}-3\chi_{,v}\!) (M/{\chi}-{M}_{,\chi}\!)\!.
\end{eqnarray}

\begin{figure}
 \includegraphics[width=9cm,height=6cm]{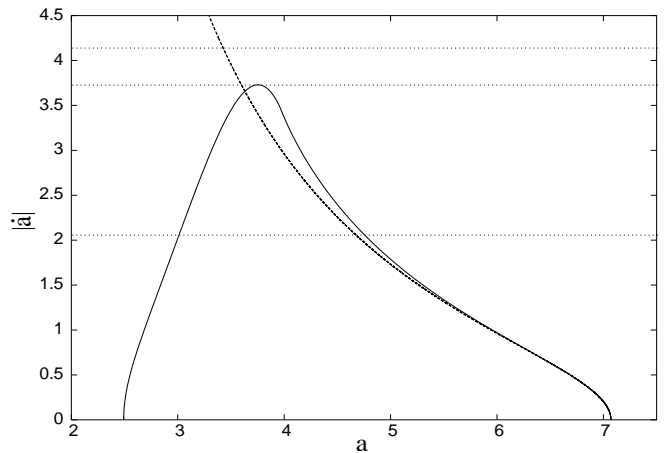}
 \caption{
The speed of collapse, $|\dot a|$, against the scale factor $a$,
for the evolution shown in Fig.~\ref{collapse_fig}, up to the
bounce. The dashed curve is for classical dynamics and
semi-classical quantum dynamics gives the solid curve. The
horizontal dotted lines correspond to different values of $R$ in
Eq.~(\ref{eq:hor}): for the upper line there is no horizon in the
quantum-corrected case, the middle line corresponds to the
threshold for a horizon, the lower line to the case of an inner
and outer horizon. \label{dota}}
\end{figure}

\begin{figure}[tbh!]
\includegraphics[width=7cm,height=7.5cm]{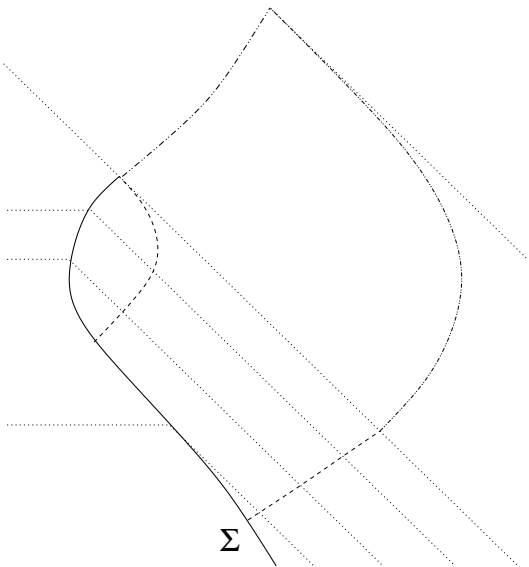}
 \caption{Eddington-Finkelstein diagram of the collapse, with boundary
 $\Sigma$. Dotted lines
 show constant $v$
 (outside) and constant $t$ (inside).
 Quantum modifications imply
 a bounce of the collapsing field, which for large enough mass is covered
 by an inner and outer evaporating horizon (dashed). A single matching
 suffices only until the inner horizon disappears. The dot-dash
 curves correspond to the subsequent evolution which is not
 determined in our model.
\label{Horizon}}
\end{figure}

The exterior region can contain trapped surfaces when the
condition $2M(v,\chi)=\chi$ is satisfied. Evaluating this
at the matching surface, using Eqs.~(\ref{eq:match1})
and (\ref{eq:match3}), gives
\begin{equation} \label{eq:hor}
 |\dot{a}|=R^{-1} (1-R^2/4)\,.
\end{equation}
When this value is reached, a dynamical horizon~\cite{DynHor}
intersects the matching surface. This always occurs classically since
during the collapse $|\dot{a}|$ varies from zero to infinity. With the
modified dynamics, however, $|\dot{a}|$ is {\em bounded} throughout
the evolution, so that {\em it depends on initial values whether or
not a horizon forms} (Fig.~\ref{dota}). Moreover, after the bounce,
$\dot{a}$ grows again, so that the condition can be satisfied a second
time. This results in a picture where the bounce, replacing the
classical singularity, may be shrouded by an evaporating dynamical
horizon outside, as shown in Fig.~\ref{Horizon}. There will be a
second point where the horizon condition is satisfied since
$|\dot{a}|$ decreases between the peak of $d_j(a)$ and the bounce.

When it intersects the matching surface, the horizon is always
null, as follows from Eq.~(\ref{Ktt}). Its later behavior depends
on the details of the outer region, which can not be determined
here. Nevertheless, one can expect that both horizons will become
timelike and evaporate. Horizon evaporation in this model does not
only come from Hawking radiation, which may be included
effectively in the outside matter content, but also from
violations of energy conditions around the bounce, which may lead
to effective outgoing negative energy.

The model is not able to specify the future of the system after it
re-emerges out of the horizon. Equation~(\ref{eq:match2}) shows
that $dv/dt$ diverges if and only if $\dot{a}>0$ and the matching
surface becomes trapped. Thus, we can describe the collapse with a
single matching until a horizon disappears, at which point the
interior $t$ ceases to be a good coordinate. One has to continue
with a second matched region to analyze the future of the system,
but this is beyond the scope of our model.

The qualitative picture that emerges from our toy model is thus
the following:  \\ $\bullet$ We do obtain black holes, i.e.,
dynamical horizons, for large masses, but they contain a bounce of
the infalling matter rather than a singularity. For large mass,
violations of energy conditions are initially small and the
evaporation takes a long time, so that
there are only small deviations from classical results. \\
$\bullet$ For small enough mass however, black holes do not form;
horizons do not develop during collapse and the bounce is
uncovered. The critical threshold scale for horizon formation is
given by the turning point in the $|\dot a|$ curve. By
Eqs.~(\ref{D}) and (\ref{fq}), the critical scale is
\begin{equation}\label{crit}
a_{\rm crit} = 0.987\ell_* =0.276 \sqrt{j}\,\lpl\,.
\end{equation}
The corresponding threshold mass is $m_{\rm crit}=m_M+m_\phi(a_{\rm
crit})$, but we are unable to compute this mass because the exterior
dynamics remains undetermined.

Our estimates may be strongly influenced by the simplifications, in
particular a homogeneous interior, we are forced to impose on the
problem. However, the qualitative features should be robust, and can
provide guidance for further more general analysis.  In particular,
they mean that there could be {\em lower bounds on the masses of black
holes that form by gravitational collapse}. This could rule out
primordial black holes below the threshold mass, and thus modify
estimates of Hawking radiation effects from very small black
holes. More speculative is an extension to highly non-spherical
situations such as particle collisions.  If LQG effects can in future
be shown to encode some of the non-perturbative aspects of string
theory, then our results may have implications for the production of
black holes in colliders, as predicted in brane-world
gravity~\cite{coll}. The black hole horizon threshold would be a
multiple not of $\lpl$, but of the higher- dimensional Planck scale,
which could be as low as $O({\rm TeV})$.  This would mean that higher
collision energies are needed for horizon formation, so that black
hole production could be significantly reduced.

{\bf Acknowledgments:} RM is supported by PPARC. PS is supported
by Eberly research funds of Penn State and NSF grant PHY-00-90091.

\end{document}